\documentclass[acmsmall,screen]{acmart}

\AtBeginDocument{%
  \providecommand\BibTeX{{%
    \normalfont B\kern-0.5em{\scshape i\kern-0.25em b}\kern-0.8em\TeX}}}

\setcopyright{none}
\copyrightyear{2024}
\acmYear{2024}
\acmDOI{XXXXXXX.XXXXXXX}

\usepackage{amsmath,amsfonts}
\usepackage{graphicx}
\usepackage{textcomp}
\usepackage{xcolor}
\def\BibTeX{{\rm B\kern-.05em{\sc i\kern-.025em b}\kern-.08em
    T\kern-.1667em\lower.7ex\hbox{E}\kern-.125emX}}

\usepackage[utf8]{inputenc}
\usepackage{algorithm}
\usepackage{algpseudocode}
\usepackage{lipsum} 
\usepackage{graphicx}
\usepackage{subcaption}

\usepackage{wrapfig}
\usepackage[binary-units=true]{siunitx}
\usepackage{multirow}
\usepackage{xspace}
\algrenewcommand\algorithmicindent{1.0em}%
\usepackage{rotating}
\usepackage{array}
\usepackage{wrapfig}

\makeatletter 
\newcommand*{\eg}{%
    \@ifnextchar{.}%
        {e.g}%
        {e.g.,\@\xspace}%
}
\newcommand*{\ie}{%
    \@ifnextchar{.}%
        {i.e}%
        {i.e.,\@\xspace}%
}
\newcommand*{\etal}{%
    \@ifnextchar{.}%
        {et al}%
        {et al.\@\xspace}%
}
\makeatother 

\begin{document}

\title{WannaLaugh: A Configurable Ransomware Emulator}
\subtitle{\large Learning to Mimic Malicious Storage Traces}

\author{Dionysios Diamantopoulos}
\email{did@zurich.ibm.com}
\orcid{0000-0003-2979-5946}
\affiliation{%
  \institution{IBM Research Europe}
  \streetaddress{S\"aumerstrasse 4}
  \city{R\"uschlikon}
  \state{Zurich}
  \country{Switzerland}
  \postcode{8803}
}
\author{Roman Pletka}
\email{rap@zurich.ibm.com}
\orcid{0000-0002-6162-3741}
\affiliation{%
  \institution{IBM Research Europe}
  \streetaddress{S\"aumerstrasse 4}
  \city{R\"uschlikon}
  \state{Zurich}
  \country{Switzerland}
  \postcode{8803}
}
\author{Slavisa Sarafijanovic}
\email{sla@zurich.ibm.com}
\affiliation{%
  \institution{IBM Research Europe}
  \streetaddress{S\"aumerstrasse 4}
  \city{R\"uschlikon}
  \state{Zurich}
  \country{Switzerland}
  \postcode{8803}
}

\author{A.L. Narasimha Reddy}
\email{reddy@ece.tamu.edu}
\orcid{0000-0003-4625-8819}
\affiliation{%
  \institution{Department of Electrical and Computer Engineering, Texas A\&M University, College Station}
  \city{Texas}
  \country{USA}
  \postcode{TX 77843}
}

\author{Haris Pozidis}
\email{hap@zurich.ibm.com}
\orcid{0000-0001-5084-6651}
\affiliation{%
  \institution{IBM Research Europe}
  \streetaddress{S\"aumerstrasse 4}
  \city{R\"uschlikon}
  \state{Zurich}
  \country{Switzerland}
  \postcode{8803}
}

\renewcommand{\shortauthors}{}

\begin{abstract}
\begin{large}
Ransomware, a fearsome and rapidly evolving cybersecurity threat, continues to inflict severe consequences on individuals and organizations worldwide. Traditional detection methods, reliant on static signatures and application behavioral patterns, are challenged by the dynamic nature of these threats. This paper introduces three primary contributions to address this challenge. First, we introduce a ransomware emulator. This tool is designed to safely mimic ransomware attacks without causing actual harm or spreading malware, making it a unique solution for studying ransomware behavior. Second, we demonstrate how we use this emulator to create storage I/O traces. These traces are then utilized to train machine-learning models. Our results show that these models are effective in detecting ransomware, highlighting the practical application of our emulator in developing responsible cybersecurity tools. Third, we show how our emulator can be used to mimic the I/O behavior of existing ransomware thereby enabling safe trace collection. Both the emulator and its application represent significant steps forward in ransomware detection in the era of machine-learning-driven cybersecurity.
\end{large}
\end{abstract}




\maketitle

\section{Introduction}

In today's cybersecurity landscape, ransomware stands as one of the most potent threats, capable of disrupting businesses and holding sensitive data hostage~\cite{xforce2023}. Traditional detection methods, based on static signatures and known patterns, struggle to adapt to the rapidly evolving tactics of cybercriminals. 
While monitoring file accesses, operating system activities or network traffic helps to reduce risks from cyberattacks~\cite{mcintosh:2021:csur}, these measures may not be enough as hackers gain internal access through social engineering techniques.
As new strains of ransomware continue to emerge, it is crucial to devise innovative solutions to combat these threats.
More advanced device-level behavioral fingerprinting using kernel events to detect anomalous patterns~\cite{sanchez2021fingerprinting, huertas2023} can be leveraged, a combination of various strategies combining OS, file-system, network, and storage information holistically is typically the most successful to efficiently address the risks.

Ransomware operates by encrypting files on a storage medium, e.g., Solid State Drive (SSD), often undetected until a ransom demand is delivered. Despite the fact that ransomware attacks stored data, their detection within storage systems has barely been studied. Nevertheless, the process leaves a unique footprint in the form of Input/Output (IO) operations - the "ransomware workload". Rudimentary work has shown that the IO operations reveal a distinct pattern that diverges significantly from non-malicious IO traces when observed~\cite{hirano:2019:8939214}. This distinctiveness arises due to the characteristic of aggressive file encryption, which leads to extensive modifications to storage data, as well as the specific access patterns that are not typical in benign workloads. We believe that computational storage devices (CSD) can be leveraged to extract feature information at line speed without impact on the host IO performance~\cite{HeydariGorji2022csd}.

Extending ransomware detection to storage systems has many advantages: (1) data extraction can be performed in parallel using a large number of CSDs, (2) the feature extraction and inference can be executed directly in the storage system stack, and (3) all these tasks can be executed without impacting the host IO traffic. At the same time, periodic snapshots in storage systems can be leveraged to mitigate an ongoing attack at low overhead.

Specifically, our paper's primary focus and contribution is to utilize these unique behavioral differences to create IO traces for training machine learning (ML) models capable of detecting ransomware attacks. Our contributions are:

\begin{itemize}
\item We introduce a state-of-the-art, highly configurable ransomware emulator designed to mimic the activity of real ransomware, without causing harm or spreading contagiously. 
\item The ransomware emulator can be tailored by a wide range of configurations, allowing users to adjust its behavior to closely resemble popular ransomware strains or create entirely new traces that represent potential future threats.
\item We provide an efficient algorithm to mimic existing ransomware behavior or foresee and emulate potential future malicious behaviors with our emulator and evaluate its performance to facilitate the development of appropriate countermeasures.
\end{itemize}

\begin{figure}[t]
  \centering
  \includegraphics[width=0.7\linewidth]{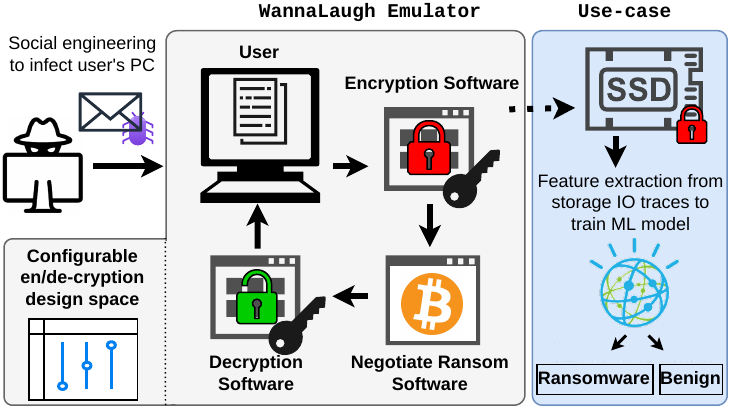}
  \caption{Life-cycle of Ransomware and positioning of WannaLaugh emulator as a use-case enabler.}  
  \label{fig:plot_composability}
\end{figure}

Our novel ransomware emulator provides numerous advantages over traditional research methods. Generating realistic, synthetic ransomware traces in a non-destructive manner, enables the collection of valuable data without exposing systems to genuine threats. Its flexibility empowers security researchers to develop ML models that are better equipped to tackle both, known and emerging ransomware variants.
The emulator can also be deployed in real-world scenarios for online ML training, as it is non-malicious and allows for the decryption of encrypted data.
Furthermore, the emulator's ethical and non-harmful nature ensures its responsible use in both research and real-world environments.

The WannaLaugh emulator, unlike a simulator, meticulously replicates ransomware behavior in realistic setups while maintaining non-malicious characteristics. This distinction enables the emulator to provide a comprehensive and accurate representation of ransomware activity, ensuring a secure research environment. WannaLaugh is designed with ethical considerations in mind and cannot be used for malicious purposes. It lacks the ability to exploit vulnerabilities or deliver dropping code, which prevents it from being contagious or spreading additional malware. Instead, WannaLaugh focuses solely on emulating encryption behavior, ensuring that it remains a safe tool for security research with no risk to systems or users.
In this article, we explore the inner workings of this ransomware emulator, examine its wide-ranging applications, and discuss the implications for ransomware detection in the future.

\section{Related Work}
\subsection{Background and Motivation}

\textbf{Understanding ransomware and benign workloads:} Our work relies on a foundational understanding of what constitutes ransomware and benign workloads. A ransomware workload refers to the specific pattern of activity exhibited by ransomware as it seeks to encrypt files on a system undetected. This behavior results in a distinctive IO trace that differs significantly from normal or 'benign' workloads. A 'benign' workload, in our context, is the inverse of a ransomware workload. It includes a broad spectrum of typical user and application activities and system operations.

\textbf{Ransomware detection techniques:} To gain the benefits of a safe ransomware emulator for generating IO traces, it is essential to understand state-of-the-art ransomware detection techniques. Researchers have proposed various detection techniques, including signature-based, behavior-based, and ML-based approaches~\cite{mohan:2020, 10.1007/978-3-642-41284-4_8, d44e2dcbbff84b93ae5e27083d8f78a2, app12010172, mcintosh:2021:csur, saracino:2018:madam, PMID:36123364, chen:2017}. While signature-based methods can detect known strains, they fail against new variants. Behavior-based detection offers adaptability but can suffer from false positives and may not effectively counter advanced obfuscation~\cite{fukushima:2010, han:2020}.

\textbf{The rise of ML/AI in ransomware detection:} Consequently, researchers have turned to ML and artificial intelligence (AI)~\cite{9711906, 9692402, 9720869, s22051837, M2023100529} and started exploring the opportunity to design ML models from storage traces~\cite{hirano:2019:8939214, HIRANO2022301314}. Analyzing disk IO traces offers several advantages, such as capturing highly indicative features correlated with ransomware activity and maintaining robustness against code obfuscation.

\textbf{The importance of ransomware emulation:} Developing storage-based ML ransomware detection techniques requires realistic data that closely resembles benign workloads and actual ransomware behavior. 

Standard IO techniques, such as synthetic workload generators or replaying captured IO traces, could potentially serve this purpose. However, they inherently lack the flexibility and dynamism needed to simulate the rapidly evolving ransomware threats. Synthetic workload generators, for example, often rely on a limited set of predefined patterns and may not fully capture the intricacy and randomness of real-world ransomware behavior. Similarly, replaying captured IO traces could limit the model's exposure to a fixed set of ransomware techniques, potentially making it ill-prepared for new and emerging ransomware strains.

Hence, in the face of these limitations, a ransomware emulator becomes a valuable tool. It can generate IO traces that realistically mimic real ransomware, doing so in a secure environment that allays ethical, compliance, and practical concerns associated with using actual ransomware. WannaLaugh, our safe and highly configurable ransomware emulator, aims to address these challenges. By allowing researchers to delve into various ransomware encryption techniques, it offers the potential to develop and validate ML models, evaluate their performance and robustness against advanced evasion, and explore novel features and indicators to enhance detection accuracy. 
By emulating a wide spectrum of ransomware behaviors, including potential future threats, WannaLaugh can significantly accelerate research in ML-based ransomware detection, leading to more effective defense strategies against this pervasive threat.

\subsection{The landscape of ransomware simulation tools}

The evolving landscape of ransomware in recent years has spurred the development of various simulation tools, reflecting the growing complexity and diversity of ransomware attacks. These tools, both open-source and commercial, shown in Table~\ref{tab:tool-comparison}, offer unique insights into ransomware behavior, yet vary significantly in their depth and applicability.

\begin{table}[ht]
\centering
\scriptsize
\caption{Comparison of various ransomware simulator/emulator tools}
\label{tab:tool-comparison}
\begin{tabular}{|r|l|c|c|c|l|c|c|}
\hline
\textbf{Tool} & \textbf{Encryption Alg.} & \textbf{OS} & \textbf{Configurations} & \textbf{Language} & \textbf{SloC} & \textbf{Update} & \textbf{License} \\ \hline
QuickBuck~\cite{quickbuck} & AES-CBC-256 & Windows & 1 & Go         & 375 & 5/2022   & MIT \\ \hline
leomatias~\cite{leomatias} & AES-CBC-256 & Windows & 1 & PowerShell & 224 & 3/2018 & MIT \\ \hline
PSRansom~\cite{psransom}   & AES-CBC-256 & Windows & 1 & PowerShell & 408 & 1/2024   & GPL 3.0 \\ \hline
GonnaCry~\cite{gonnacry}   & AES-CBC-256 & Linux   & 1  & Python   & 670 & 10/2020  & GPL 2.0 \\ \hline
Lawndoc~\cite{ransim}   & AES-CBC-256 & Windows   & 1  & PowerShell   & 670 & 1/2023  & MIT \\ \hline
shinolocker~\cite{shinolocker} & AES-128 & Windows & N/A & PowerShell & N/A & 2016       & Freeware \\ \hline
carbonsec~\cite{carbonsec}     & N/A     & Windows & N/A         & N/A        & N/A & 2024       & Commercial \\ \hline
spin.ai~\cite{spinai}          & N/A     & Windows & 11         & N/A        & N/A & 2024       & Commercial \\ \hline
RanSim~\cite{ransim-knowbe4}   & N/A     & Windows & 25         & N/A        & N/A & 2024       & Commercial \\ \hline
\multirow{5}{*}{WannaLaugh} & AES 128/192/256, & \multirow{5}{*}{Windows,} & 36 Encryption Alg/s & \multirow{6}{*}{Python} & \multirow{6}{*}{4012} & \multirow{6}{*}{1/2024} & \multirow{6}{*}{Apache v2} \\
\multirow{5}{*}{(this work)} & Modes: CBC, XTS, & \multirow{5}{*}{Linux,} & 3 write options & & & & \\
& ECB, GCM, CTR, & \multirow{5}{*}{MacOS} & n-threads, t-timeout & & & & \\
& CFB, OFB, CCM, & & k-encrypt segment & & & & \\
& EAX, OCB, CTS & & l-skip segment & & & & \\
\cline{2-2}
& SALSA20, CHACHA20 & & \( 108 \times |N|\times|K| \times |L| \) & & & & \\
\hline
\end{tabular}
\end{table}

In the realm of commercial ransomware simulators, such as KnowBe4's RanSim~\cite{ransim-knowbe4}, carbonsec~\cite{carbonsec}, and spin.ai~\cite{spinai}, the primary concern lies in their 'black box' nature. These tools, while potentially useful for practical applications, are often opaque in their operation. This lack of transparency not only poses challenges for academic research, where understanding the underlying mechanisms is crucial but also raises questions for enterprises about the exact functionalities being executed within their systems. For instance, KnowBe4's RanSim~\cite{ransim-knowbe4} uses 23 fixed penetration testing scenarios (21 ransomware and 2 benign) but includes potentially dangerous practices, such as injecting code into legitimate processes. In addition, it provides only a very small set of test files. The proprietary nature of these types of tools limits the scope for customization and detailed analysis, making them ineligible for research and a nuanced understanding of ransomware tactics.

In the open-source ecosystem, existing simulators come with notable variations in capabilities. Tools like GonnaCry~\cite{gonnacry} and Lawndoc/RanSim~\cite{ransim} have made valuable contributions, offering basic encryption functionalities and implementation-specific features. GonnaCry, for instance, addresses Linux-based ransomware simulation only.  However, its limitations (\eg deprecated PyCrypto library, compatibility issues with recent Python versions, and support for text files only) and no controllable options, would have required a complete rewrite of the code.  Tools like Lawndoc/RanSim~\cite{ransim}, QuickBuck~\cite{quickbuck}, leomatias~\cite{quickbuck}, PSRansom~\cite{psransom}  provide a simple approach to file encryption and decryption and, while effective in their implementation of AES-CBC-256 encryption, are limited in their encryption algorithm diversity. They predominantly focus on a single encryption method and are tailored to the Windows OS. This lack of variety in encryption strategies restricts their utility in simulating the wide array of ransomware attacks seen in the real world. As such, these tools lack the comprehensive features necessary for simulating the full spectrum of ransomware behavior.

In contrast, WannaLaugh represents a significant advancement in ransomware simulation. Its support for a wide array of AES encryption key lengths (128, 192, 256) and eleven AES modes, along with streaming ciphers like SALSA20 and CHACHA20, places it at the forefront of ransomware research tools. The encryption capabilities are further augmented with specialized capabilities such as configurable intermittent encryption, a frequently used technique in some ransomware that utilizes partial encryption to improve efficiency and better evade detection. Additionally, the broad multi-operating system support of WannaLaugh, encompassing Windows, Linux, and MacOS, along with its distribution under the Apache v2 license, further solidifies its standing as an ideal tool for research. The Apache v2 license not only ensures transparency but also encourages community engagement, allowing researchers, developers, and cybersecurity professionals to study, understand, and extend the tool's capabilities. This feature set, together with Wannalaugh's configurability over the entire encryption process, which is discussed in the following section, starkly contrasts with the more limited functionalities of existing open-source tools.

\section{Ransomware Emulator Design}

The effectiveness of a ransomware emulator hinges on its ability to accurately replicate the complete ransomware lifecycle. Following this concept, WannaLaugh consists of three fundamental components: the encryptor, the decryptor, and the server, which manages ransom negotiations. Many ransomware variants employ unique encryption patterns, making it imperative that our emulator not only encrypts data but also demonstrates the capability to decrypt it accurately. This ensures that we have a valid and comprehensive emulation of the encryption-decryption process as it occurs in real-world scenarios. Without this, the emulator would fall short in providing a reliable platform for researchers to study the full spectrum of ransomware behavior, including potential recovery methods post-encryption.

Regarding the inclusion of the server component, it's important to consider the broader implications for cybersecurity research. The server's role extends beyond simple ransom negotiation; it offers the capability to simulate various response strategies and negotiation handshakes. This feature is particularly valuable for generating network traces, which can be used by researchers focusing on network anomaly detection as a means to identify ransomware attacks. Although this aspect of the server is not the primary focus of our current research, we recognize its significance and potential applications. Therefore, we included this component to allow the community to further explore and develop tools for network-based ransomware detection.

Furthermore, another crucial reason for incorporating the server component in our tool is to potentially emulate Command and Control (C2) setups for emerging ransomware variants that include data exfiltration techniques. This emerging trend in ransomware attacks involves not only encrypting the victim's data but also exfiltrating sensitive information to a server controlled by attackers. Before initiating the encryption process, these advanced ransomware strains may send valuable data to a server.

By including a server that accepts these exfiltrated files, our tool can simulate this sophisticated attack architecture, providing a comprehensive emulation of modern ransomware tactics. While data exfiltration is not the primary focus of this paper, its inclusion in the server component significantly enhances the tool's capability to emulate these advanced ransomware strategies. This addition allows researchers and cybersecurity professionals to test and develop defensive techniques against ransomware that not only encrypts but also steals data, representing a more severe threat to organizations.

\subsection{Configurable design space}

When developing WannaLaugh, we have carefully analyzed the file IO patterns exhibited by six real ransomware strains, namely Sodinokibi~\cite{sodinokibiencryption}, BlackBasta~\cite{Black-basta-trendmicro}, Lockbit~\cite{lockbitencryption}, Lockfile~\cite{X-Force_Lockfile}, WannaCry~\cite{wannacryencryption}, and Conti~\cite{decodingconti}. This comprehensive study of actual ransomware behavior has driven the design of the configurable options in the emulator. By incorporating feedback from observed file IO patterns, WannaLaugh aims at replicating the characteristics of real ransomware, allowing researchers to generate realistic disk IO traces that closely resemble those from genuine ransomware. How accurately WannaLaugh manages to do that, is shown later in our analysis at Subsection~\ref{ssec:mimic_traces}.

This data-driven approach enables WannaLaugh to provide the flexibility needed for exploring not only existing ransomware variants but also future threats. By altering the parameters of our emulator, we can further generate entirely new IO traces that represent hypothetical ransomware variants. These new traces are valuable for "stress-testing" existing ransomware detection techniques and encouraging the development of more robust, generalizable techniques capable of detecting unseen ransomware types. By doing so, we ensure that our approach is not limited to generating features about existing ransomware but can also contribute to the defense preparation against future, unknown threats.

Below we provide an explanation of the configurable options and their significance in the design and implementation of WannaLaugh.

\subsubsection{Workload type}
WannaLaugh offers a comprehensive suite of workload types, categorized into three main areas: "Ransomware", "Benign", and "Mixed". These categories are designed to cover a broad spectrum of IO patterns, thereby enabling WannaLaugh to generate a diverse range of traces for in-depth ransomware research and simulation.

\textbf{Ransomware Workloads:} Detailed later in this subsection, the ransomware workload category in WannaLaugh is equipped with extensive design options. This category is meticulously crafted to emulate various ransomware behaviors, allowing researchers to study a wide array of ransomware attack patterns. 

\textbf{Benign Workloads:} In benign mode, WannaLaugh emulates a rich array of non-malicious workload types, categorized into three distinct groups: \textit{file conversions, Filebench workloads}~\cite{DBLP:journals/usenix-login/TarasovZS16}, and \textit{compression workloads}.

\textit{File Conversions:} This category encompasses a range of everyday file operations and format conversions, vital for establishing a "benign" baseline. Included conversions cover a wide spectrum, such as pdf-to-doc, ps-to-pdf, doc-to-pdf, xlsx-to-csv, json-to-csv, txt-to-docx, html-to-pdf, png-to-jpeg, jpg-to-png, gif-to-bmp, and gz-to-zip. These activities mirror typical user behavior, generating IO traces that reflect a foundation of common, non-malicious operations.

\textit{Filebench Workloads:} The integration of Filebench, a versatile file system benchmark tool, into WannaLaugh adds a new dimension to the benign mode.  It is integrated as an external tool due to its C-based codebase. WannaLaugh provides parameters to control the number of files, their sizes, and the number of threads for different Filebench workloads (called personas in Filebench terminology), such as OLTP (Online Transaction Processing), web server, file server, video server, web proxy, and mail server. Recognizing that default Filebench data (random or zero-populated) may not accurately represent realistic IO patterns - especially when entropy-related information is extracted, we have modified Filebench to utilize user-provided data. This allows, for example, the video server persona to serve specific files from a video corpus, thereby generating more authentic IO traces.

\textit{Compression Workloads:} For compression tasks, WannaLaugh supports several algorithms and archive programs, all of which are integrated directly into the Python codebase, eliminating the need for external tools. This category includes ZIP, LZMA, BZ2, GZ, 7z, Zstd, and LZ4. Unique to this workload is the ability to control the number of threads, the compression level (when supported by the algorithm), and the number of files per archive output. This feature offers significant flexibility and precision in simulating various compression scenarios.

\textbf{Mixed Workloads:} The mixed mode in WannaLaugh allows for a combination of ransomware and benign activities, based on a user-defined mixing rate. This mode is particularly valuable as it provides a more realistic representation of typical use cases, wherein the user's workload is running in parallel with an active ransomware infection. By incorporating the mixed mode, researchers can gain deeper insights into the complex interplay between malicious and non-malicious activities, ultimately enhancing the robustness of their detection models.

\subsubsection{Directory Selection and File Ordering} The emulator allows choosing one or more target directories, enabling to focus on particular sets of files. Further, it provides various file ordering options (\ie by name, modification time, creation time, file size, or, randomly). This flexibility allows users to simulate ransomware that employs different strategies for selecting and encrypting target files. WannaLaugh also supports filtering files by extensions or wildcards, to customize which files will be encrypted and which ones will be skipped. This feature helps to simulate ransomware that targets specific file types or employs more complex selection patterns (\eg skipping OS-specific files like $.so, .dll, .exe, .msi$, etc.). 

\subsubsection{Encryption Algorithms and Methods} 
Ransomware can use different encryption methods, such as symmetric, client asymmetric, server asymmetric, or a hybrid approach. Symmetric encryption is a fast encryption method, but it is vulnerable to decryption by researchers because the keys used for encryption and decryption are stored on disk in an unencrypted form. Client asymmetric encryption is a slow process that requires an internet connection for the ransomware to communicate with the server. The ransomware generates a pair of RSA keys, encrypts the files with the public key, and sends the private key to the server. Server asymmetric encryption is a scheme where the server generates a key pair, and the public key is hardcoded on the ransomware. Each file is encrypted with the server's public key, and only the server's private key can recover the files. However, the server would need to send the private key to the client for decryption, which is impractical and insecure. To address this, the most widely used technique by ransomware is implemented in WannaLaugh and uses a combination of symmetric and asymmetric encryption. First, it generates an asymmetric key pair, encrypting its private key with the server's public key, and then it encrypts the files using a symmetric encryption algorithm such as AES. All symmetric keys are encrypted with the generated public key. This approach is fast and requires no Internet connection for the targeted system.

The encryption capabilities of WannaLaugh are expanded to cover most of the commonly used algorithms. Building upon this, WannaLaugh supports AES with different key lengths: 128, 192, and 256 bits. Complementary it uniquely offers eleven modes of AES encryption, each providing varied operational characteristics. These modes are 1) CBC (Cipher Block Chaining), 2) ECB (Electronic Codebook), 3) GCM (Galois/Counter Mode), 4) CTR (Counter), 5) CFB (Cipher Feedback), 6) OFB (Output Feedback), 7) CCM (Counter with CBC-MAC), 8) EAX (Authenticated Encryption with Associated Data), 9) OCB (Offset Codebook), 10) CTS (Cipher Text Stealing), and 11) XTS (XEX-based Tweaked CodeBook with CipherText Stealing). This selection of modes, encompassing a broad spectrum of cryptographic functionalities, allows for varied configurations of tags and nonces, significantly influencing the I/O traces generated during encryption. This almost exhaustive selection of modes offers a distinct advantage over other tools that predominantly rely on a single mode, such as AES-CBC-256.

Furthermore, WannaLaugh also supports advanced streaming ciphers such as SALSA20 and CHACHA20, which are increasingly utilized in modern ransomware encryption scenarios. These streaming ciphers are especially efficient for partial encryption strategies, as they result in an encrypted output size identical to the original, without the need for additional elements like nonce or tags. Also, WannaLaugh supports a custom SHUFFLE algorithm that randomly swaps segments, instead of encrypting the file.

\subsubsection{Encryption Content Methods} 

The emulator allows one to choose among a diverse range of encryption methods. First, it can encrypt the entire content of a file, mirroring the common approach employed in many real-world ransomware attacks. In addition, the emulator caters to more targeted encryption strategies including highly configurable intermittent encryption methods. It allows for encrypting only the first or last bytes of a file, aligning with ransomware that partially encrypts files to speed up the process while still rendering the files unusable. Furthermore, WannaLaugh offers a unique segment-based encryption method. This mode provides the flexibility to encrypt specific segments of a file, determined by user-defined parameters for the size of the segments to encrypt and the size of the segments to skip. Such methods are employed by recent ransomware to evade detection or to reduce encryption time.

\subsubsection{Encryption Write Methods} The emulator provides three distinct methods for writing encrypted files: (1) overwriting the original file, (2) shredding the original and then copying the encrypted data to a new file, or (3) copying the encrypted data to a new file and then shredding the original. These options allow the exploration of the impact of different ransomware writing methods correlated to different block-level IO patterns on detection and response strategies.

\subsubsection{Delay Modes and Timeouts} To simulate different encryption speeds and time intervals, WannaLaugh supports three delay modes: no delay, static delay, and random delay. Users can also define a range of dynamic delays between file encryption, the number of files to encrypt continuously before a delay, or a timeout that allows users to specify a time limit for the encryption process.  Upon a timeout, the current file is skipped, simulating the ransomware behavior of evading detection by limiting its activity within a certain time range.

\subsubsection{Custom File Extensions} WannaLaugh enables users to define a custom file extension for encrypted files, allowing them to resemble the behavior of specific ransomware strains, or create new variants that use unique file extensions.

\subsubsection{Multi-threaded File Encryption using Ray} A significant performance optimization we introduced to the WannaLaugh ransomware emulator is multi-threaded encryption. This was motivated by the observation that many real-world servers possess a substantial amount of hardware-level parallelism, which could be exploited by ransomware to significantly accelerate their encryption process. We achieved this enhancement by integrating Ray~\cite{222605}, a high-performance distributed execution framework, into our ransomware emulator.

Ray was specifically selected due to its data layout for zero-copy serialization (Apache Arrow), in-memory object store (Plasma), and its proficiency at bypassing the Global Interpreter Lock (GIL) problem prevalent in Python. These characteristics facilitate efficient file encryption with zero-copy operations, significantly reducing the I/O overhead during the encryption process. We have implemented user-configurable thread settings for the encryption process. Users can specify the desired number of threads to utilize, allowing them to simulate different levels of system resource exploitation by ransomware. For more dynamic emulation, WannaLaugh also offers an automatic core adjustment feature. With this, the emulator can continually monitor and adjust the number of cores it utilizes during runtime, aiming to keep the system load under a user-specified percentage. This capability is particularly valuable for emulating sophisticated ransomware that deliberately moderates its resource usage to evade detection by users or security software.


\subsection{Implementation Aspects}
\subsubsection{Code-base}
WannaLaugh is developed in Python3, ensuring cross-platform compatibility and ease of use. For encryption/decryption, it relies on the Cryptodome package~\cite{cryptodome}. All three components, the encryptor, the decryptor, and the ransom negotiation server, support loggers to track detailed run-time or debug information. It also provides users with two convenient usage modes. For those preferring a direct approach, the emulator can be seamlessly operated through Python when installed. This method is ideal for users who wish to engage with the emulator in a more hands-on, script-based environment. This mode can also be used by developers who wish to directly use the API of WannaLaugh into their source code. Alternatively, users can leverage the \textit{PyInstaller} package to generate standalone binaries. This approach simplifies deployment across various platforms, eliminating the need for installing multiple libraries or managing several files. With just a single executable file, WannaLaugh can be effectively utilized on Windows, Linux, and macOS systems, significantly enhancing its accessibility and ease of use.

\subsubsection{Graphical User Interface (GUI)}

\begin{wrapfigure}{r}{0.55\textwidth}
  \begin{center}
    \includegraphics[width=0.50\textwidth]{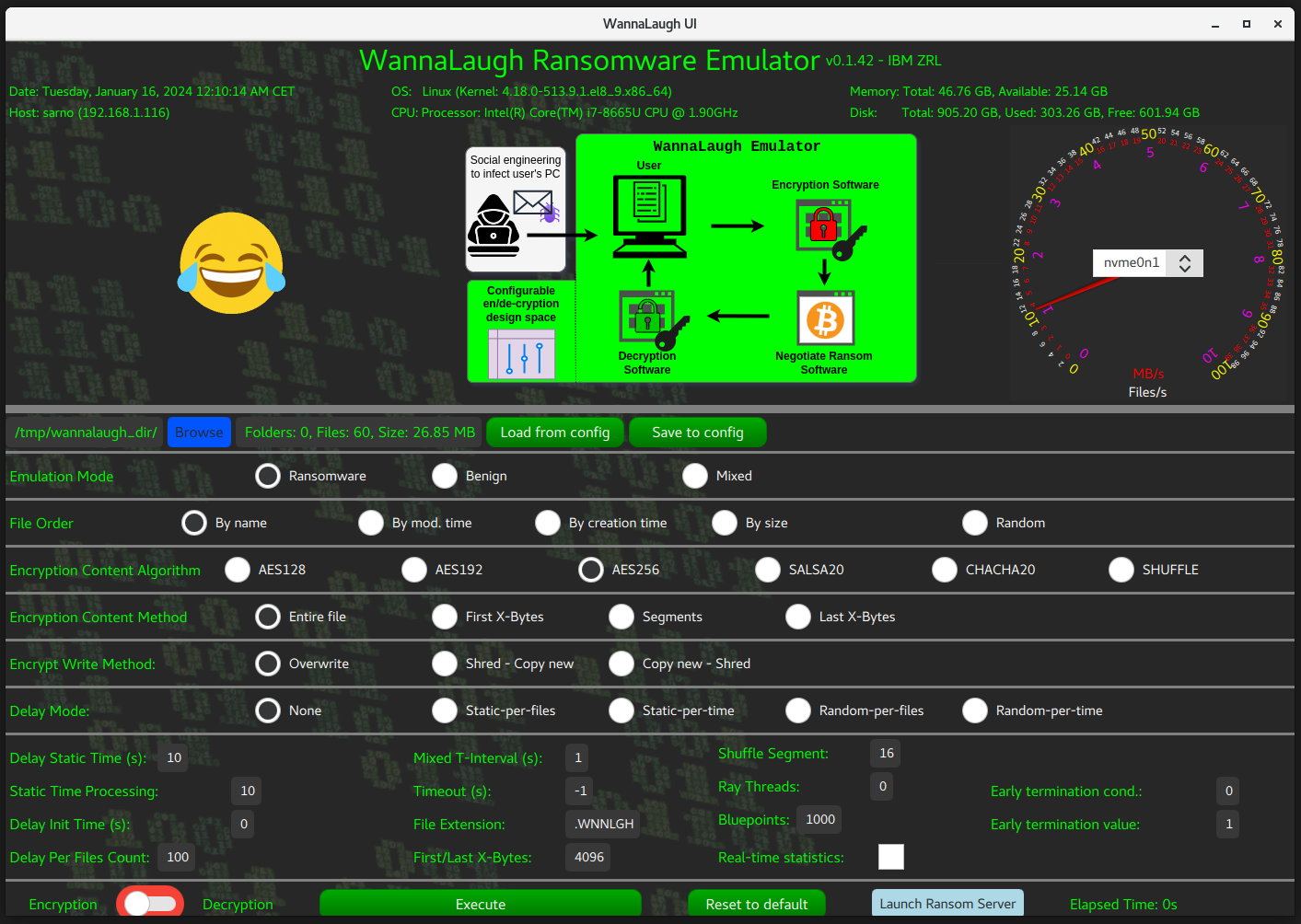}
  \end{center}
  \caption{WannaLaugh's interactive GUI.}
  \label{fig:gui}
\end{wrapfigure}

Furthermore, the inclusion of a user-friendly GUI is a deliberate design choice to enhance the accessibility and usability of WannaLaugh. A GUI allows researchers and users to interact with the emulator more intuitively, avoiding the complexities and limitations of command-line interfaces. This interactivity is vital for facilitating a broader range of experiments and for allowing users with varying levels of technical expertise to engage effectively with the emulator. In the context of ransomware detection and analysis, the GUI provides a practical and efficient way to configure, control, and observe the emulator's behavior, thus contributing significantly to the tool's educational and research utility.

The GUI of WannaLaugh is developed using QML (Qt Modeling Language)~\cite{qml}, a choice driven by the need for cross-platform compatibility, which is often a challenging aspect for GUIs. By leveraging QML, we ensure that WannaLaugh's user interface is operable and consistent across different operating systems, including Windows, Linux, and macOS. Furthermore, the architecture of WannaLaugh's GUI is meticulously designed to optimize performance. One key aspect of this architecture is the separation of the rendering thread from the emulator's core processing. This separation means that the emulator's operations are not hindered by the GUI; it can run independently at its normal performance level. This design decision is crucial for ensuring that the emulator maintains high efficiency and reliability. Additionally, by isolating the GUI in this manner, we mitigate the risk of GUI-related issues impacting the emulator's functionality. In scenarios where the GUI may encounter problems or slowdowns, the emulator itself remains unaffected and continues to operate smoothly.

\section{Application of WannaLaugh in Ransomware Detection from Storage}

Building on the detailed description of the WannaLaugh emulator provided in the previous section, this new section transitions into a practical application of WannaLaugh. Specifically, we demonstrate how such a tool can be useful in advancing ransomware detection methods through machine learning (ML) models.

\begin{figure}[h]
  \includegraphics[width=1.0\linewidth]{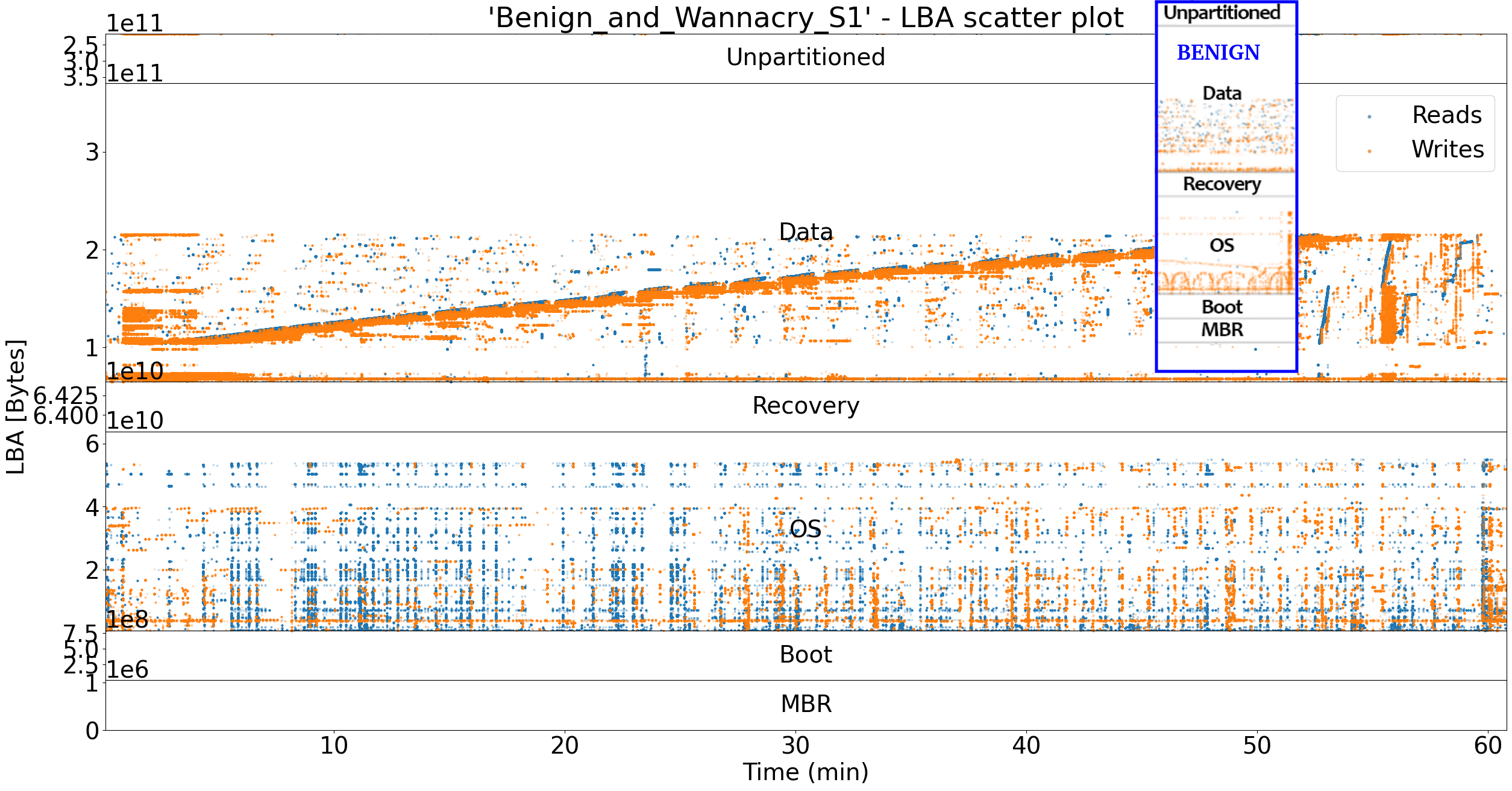}
  \caption{LBA scatter plot for WannaCry Ransomware.}  
  \label{fig:lba_scatter}
\end{figure}

In this study, we adopt a feature extraction approach based on information derived solely from IO operations inspired by Hirano \etal ~\cite{hirano:2019:8939214}. 
They trained three different ML models, with their best-performing model achieving an accuracy of 98\%. Their methodology involved collecting I/O traces from storage systems running seven ransomware and five benign workloads using Wayback-Visor, a hypervisor situated between the hardware and the operating system. They monitored the traces for several minutes, starting immediately after the workload was initiated. From these traces, they calculated five features over different window sizes (T\_window) ranging from one to 25 seconds, with a one-second shift between windows until the end of the trace time was reached. These features included the average normalized Shannon entropy of writes (H\_write(t)) and the variance of the logical block address (LBA). The windowed features were then employed to train and evaluate Random Forest, KNN, and SVM models, with the best performance achieved by the Random Forest model using a T\_window of 15 seconds~\cite{hirano:2019:8939214}.

To elucidate the concept of using one of the storage-related metrics, e.g. LBA variance, as a feature for machine learning models in ransomware detection, we present a specific plot in our study. This visualization aims to provide a simplified example of how ransomware activities can be distinguished from benign operations based on IO trace patterns, particularly focusing on the LBA variance. Specifically, Fig.~\ref{fig:lba_scatter} displays a scatter graph with the timeline in minutes of an actual Windows operating system session on the x-axis, while the y-axis represents the LBA. The data points in the plot are color-coded, with blue indicating read operations and orange signifying write operations. This visualization is derived from running a specific ransomware, 'Blackbasta,' on a Windows OS environment. Note that the pattern shown is very specific to this ransomware and the underlying NTFS file system. We have studied many other ransomware samples as well and can confirm the existence of distinguishing typical patterns in other ransomware and file system types.

Notably, the LBA space on the y-axis is categorized into several segments: the Master Boot Record (MBR) of the disk, the Boot partition, the OS partition, the Windows Recovery partition, and the Data partition. This categorization helps in understanding the distribution and intensity of read and write operations across different sections of the disk. A key observation from this plot is the distinctive pattern of read and write operations in the Data partition when ransomware is active. There is a discernible trend of intense read and write activities, starting from lower LBAs and extending towards higher LBAs in a nearly linear fashion. This pattern suggests that the ransomware sequentially accesses and encrypts files, which is a crucial insight. In contrast, the OS partition does not exhibit such intense or structured activity.

For comparative analysis, a smaller inset picture within the main plot shows a similar graph, but for a benign workload. This includes normal OS activities and browsing activities with 20 tabs of Firefox. The comparison clearly highlights the difference in read-write patterns in the Data partition under benign conditions, where the operations appear more spatially distributed and lack the structure, and sequential pattern observed in the ransomware scenario. 

Observing the effectiveness of features like LBA variance, as demonstrated in the previous analysis, we recognize the potential of similarly insightful trends in other characteristics such as entropy, read rate, and write rate. These features, effectively employed by Hirano et al. in their machine-learning models, have shown substantial accuracy in ransomware detection.

\begin{figure}[h]
  \hspace{-0.107cm}
  \centering
  \includegraphics[width=1.0\linewidth]{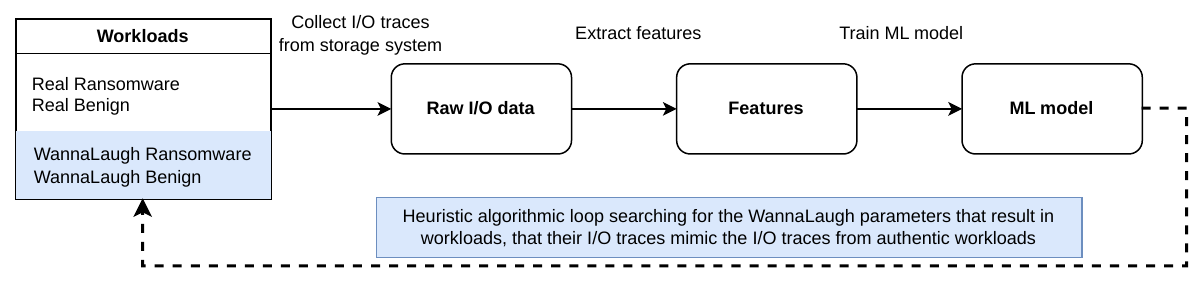}
  \caption{Overview of our ML flow and data lifetime.}
  \label{fig:mlflow}
\end{figure}

Building on this foundational work, we have extended their approach by not only incorporating similar features but also introducing new tools and techniques tailored to our unique research context. Specifically, we generate traces from six real ransomware samples, one benign workload alongside those from the WannaLaugh emulator, all of them running in a sandboxed virtual machine (VM).

Our approach differs from Hirano~\etal~\cite{hirano:2019:8939214} in that, while they focused on analyzing traces from actual ransomware and benign applications, our research also includes traces generated by the WannaLaugh emulator. This inclusion serves a dual purpose: firstly, it allows us to validate the emulator's efficacy in producing realistic ransomware-like IO patterns; secondly, it aids in developing a comprehensive methodology to ensure these emulator-generated traces align closely with authentic ransomware features. Figure~\ref{fig:mlflow} highlights the overview of our ML flow, where with blue color we annotate the the differences of this work from~\cite{hirano:2019:8939214}. The dashed line shows the exploration mechanism we introduce to find the WannaLaugh parameters that result in
workloads, their I/O traces mimic the I/O traces from authentic workloads. This path is further detailed in subsection~\ref{wan_par_sel}.

\begin{figure}[h]
  \centering
  \includegraphics[width=0.6\linewidth]{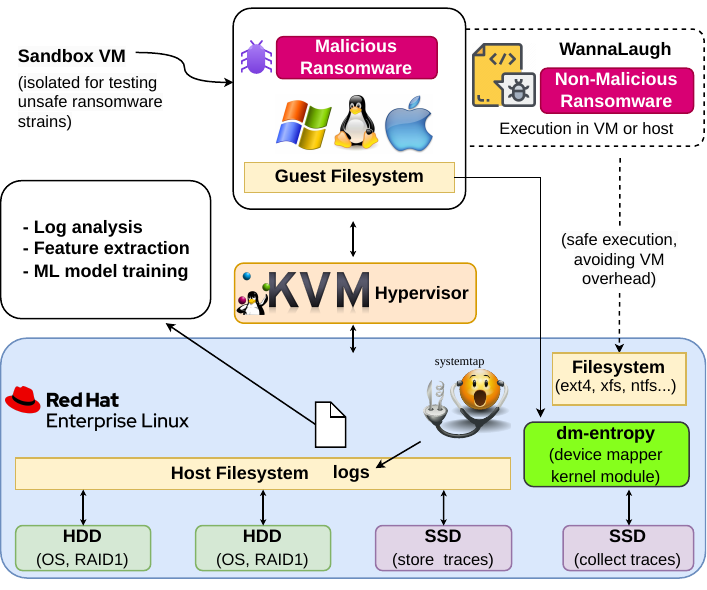}
  \caption{Test Environment for Feature Extraction.} 
  \label{fig:plot_test_env}
\end{figure}

The test environment, as depicted in Fig.~\ref{fig:plot_test_env}, is based on RHEL 9.1 using the KVM hypervisor. This setup provides the necessary isolation when running potentially malicious code and ensures the integrity of our research. The guest VMs, populated with a subset of the Govdoc1 dataset used as decoy~\cite{garfinkel:2009}, are each assigned a dedicated NVMe SSD. Utilizing the device mapper kernel module \emph{dm-entropy} we developed, we extract entropy information on writes and implement a SystemTap~\cite{systemtap:2005:ols} hook to meticulously track all IO operations on these block devices.

The binary logarithm that is needed to calculate the entropy is approximated by using the built-in CPU instruction \texttt{CLZ} which counts the leading zeros, hence resulting in extremely low overhead.
From the collected IO traces which include a timestamp and the Shannon entropy of writes, the same features are extracted offline over a given time window as in~\cite{hirano:2019:8939214}, namely the mean Shannon entropy, the read and write throughput, and the variance of the logical block address (LBA) reads and writes.  For simplicity, we used a window size of {\SI{2}{\second}} with no overlap of windows.  The collected traces are then used to train supervised ML models (\eg Random Forest, XG Boost, and DNN) which we evaluated using 5-fold cross-validation. 

\begin{figure}[h]
  \hspace{-0.107cm}
  \centering
  \includegraphics[width=0.6\linewidth]{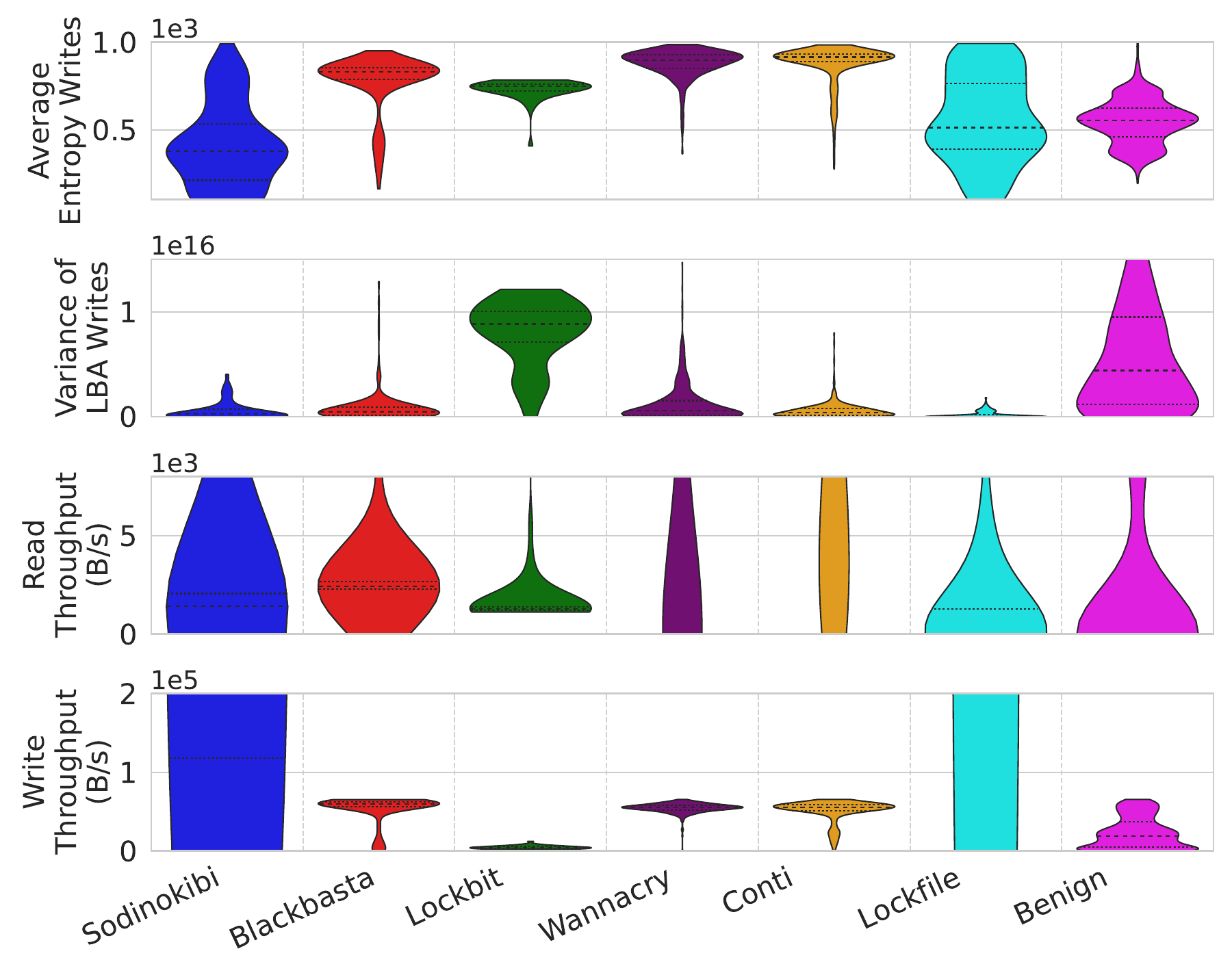}
  \caption{Variation of features for ransomware strains.}  
  \label{fig:violin_plot_features}
\end{figure}

Fig.~\ref{fig:violin_plot_features} shows violin plots that provide a visual representation of the distribution of Shannon entropy and variance of LBA writes of the real ransomware strains and a benign workload performing file conversions.
Each violin plot illustrates the probability density of the data at different values, with the width of the plot indicating the density of the data.
The plots reveal significant variations among the features within and across individual ransomware strains. WannaLaugh's ability to generate diverse feature shapes on-demand is essential, as it supplies the variation required for ML models to effectively learn distinct ransomware behaviors.

\subsection{WannaLaugh Parameter Selection}
\label{wan_par_sel}

The goal of the emulator is to facilitate the development and evaluation of ransomware detection techniques by providing a realistic, yet safe environment for researchers to study file access patterns and encryption behaviors. As already discussed, the WannaLaugh emulator is highly configurable, allowing security researchers to explore a wide range of ransomware behaviors. However,
selecting the appropriate configuration options to guide the generation of traces to match specific ransomware is a challenging task, considering the extensive design space.
We present an algorithm for selecting WannaLaugh parameters that result in traces either resembling or differing from known ransomware samples to address this issue.
The algorithm searches through the design space to guide the selection of configuration options, ultimately generating IO traces that closely resemble or diverge from the target ransomware or workload as desired.

We initially employed a random walk (RW) algorithm as a fundamental search strategy to explore the design space by iteratively selecting random configuration options and evaluating their fitness based on the closeness of the generated trace to the target ransomware. Despite its simplicity, the RW algorithm did not converge as rapidly as desired. Consequently, we explored alternative optimization techniques, such as greedy genetic algorithms (GGA)~\cite{katoch2021review}, simulated annealing (SA)~\cite{delahaye2019simulated}, and the Non-dominated Sorting Genetic Algorithm II (NSGA-II)~\cite{996017}, to guide the search more efficiently. The parameters for these algorithms are shown in Table~\ref{tab:algorithm_parameters}.

Upon comparing the performance of these optimization techniques,
NSGA-II demonstrated the best convergence, such that the generated traces either resemble known ransomware traces or do not match any of the studied ransomware traces. This dual-purpose approach enables the evaluation of ransomware detection tools by stressing them with new, unknown ransomware samples.
NSGA-II in conjunction with an ML classification model effectively optimizes the WannaLaugh parameter configurations.  The decision to showcase only the NSGA-II in this study, which is presented in Algorithm~\ref{alg:alg1}, stems from its superior convergence characteristics.

\begin{algorithm}[tb]
\caption{Guiding WannaLaugh to resemble (or not) malicious traces using NSGA-II and ML}
\begin{algorithmic}[1]
\algrenewcommand{\algorithmiccomment}[1]{\hskip1em$\rightarrow$ #1}
\scriptsize
\Procedure{OptimizeWannaLaughParameters}{}
\State Initialize population $P_0$ with random WannaLaugh parameter configurations
\State Evaluate $P_0$ by generating traces $T_{generated}$, classifying using $\Call{Classify}{T_{generated}}$, and calculating cost $c = \Call{Cost}{F1\_score}$
\State $t \gets 0$

\While{termination condition not met}
    \State Generate offspring population $Q_t$ using (crossover, mutation) on $P_t$
    \State Evaluate $Q_t$ by generating traces $T_{generated}$, classifying using $\Call{Classify}{T_{generated}}$, and fitness cost $c = \Call{Cost}{F1\_score}$
    \State Combine parent and offspring populations: $R_t \gets P_t \cup Q_t$
    \State Perform non-dominated sorting and crowding distance calculation on $R_t$
    \State Select best from $R_t$ for $P_{t+1}$ based on rank \& crowding distance (elitism)
    \State $t \gets t + 1$
\EndWhile

\State \textbf{return} optimized WannaLaugh parameter configurations in $P_t$
\EndProcedure

\Function{Classify}{T\_generated}
    \State Load pre-trained ML model and Perform inference using $T_{generated}$ as input
    \State \textbf{return} predicted ransomware class and $F1\_score$
\EndFunction

\Function{Cost}{$F1\_score$, \textit{optimization\_goal}}
    \State if \textit{optimization\_goal} is "resemble" \textbf{return} cost $c = 100 - (F1\_score * 100)$ 
    \State else \textbf{return} cost $c = F1\_score * 100$  // new traces not resembling any strains
\EndFunction
\end{algorithmic}
\label{alg:alg1}
\end{algorithm}

\begin{table}[h]
\centering
\footnotesize
\caption{Parameters of the selected algorithms}
\label{tab:algorithm_parameters}
\begin{tabular}{ll||ll}
\hline
\multicolumn{2}{l||}{\textbf{Random Walk}} & \multicolumn{2}{l}{\textbf{Simulated Annealing}} \\
\hline
Step size & 1 & Initial temperature & 1000 \\
Parallel executions & 100 & Cooling rate & 0.95 \\
Max steps & 1000 & Iterations & 1000 \\
\hline
\hline
\multicolumn{2}{l||}{\textbf{Greedy Genetic Algorithm}} & \multicolumn{2}{l}{\textbf{NSGA-II}} \\
\hline
Population size & 100 & Population size & 100 \\
Generations & 1000 & Generations & 1000 \\
Crossover probability & 0.8 & Crossover probability & 0.8 \\
Mutation probability & 0.1 &  Mutation probability & 0.1 \\
 & & Elitism percentage & 10\%-30\% \\
\hline
\end{tabular}
\end{table}

The algorithm commences by initializing a population of random WannaLaugh parameters, which are then evaluated by generating traces and classifying them using the pre-trained ML model's classification function. 
The cost function is determined based on the optimization goal, resemblance or non-resemblance of ransomware traces, and uses the F1 score from classification.
The inclusion of the ML classification model within the optimization process ensures the effectiveness of the algorithm in identifying traces that either closely resemble or diverge from known samples.
During the iterative process, offspring populations are generated using crossover and mutation operations. These offspring populations are evaluated, combined with the parent populations, and undergo non-dominated sorting and crowding distance calculations. The best solutions, based on rank and crowding distance, are selected for the next iteration. The algorithm continues until the termination condition is met, yielding optimized WannaLaugh parameters that achieve the desired optimization goal.

\section{Case Studies and Experiments}

In this section, we present the evaluation results of the multi-threaded encryption feature of WannaLaugh, along with the convergence performance of the four optimization algorithms and the ability of WannaLaugh to mimic traces.

\subsection{Evaluation of Multi-threaded Encryption}

Given that modern NVMe SSDs offer substantial bandwidth that single-threaded operations cannot fully utilize, it is critical to understand the potential impact of ransomware that leverages hardware-level parallelism for file encryption. This could significantly expedite an attack by enabling the ransomware to approach the SSD's bandwidth limits, hence encrypting files orders of magnitude faster. Consequently, we have conducted an evaluation of WannaLaugh's multi-threaded encryption feature to highlight its potential implications.

To evaluate the performance of multi-threaded encryption, an experiment was conducted where WannaLaugh was tasked to encrypt the first 60 folders from the Govdoc1 dataset~\cite{garfinkel:2009} on a Linux server equipped with 40 physical cores (each with 2 threads-per-core). The folders comprised 59,272 files (44.7GB), with 58,213 being eligible for encryption (e.g. log files are omitted).

\begin{wrapfigure}{r}{0.65\textwidth}
  \centering
    \includegraphics[width=0.6\textwidth]{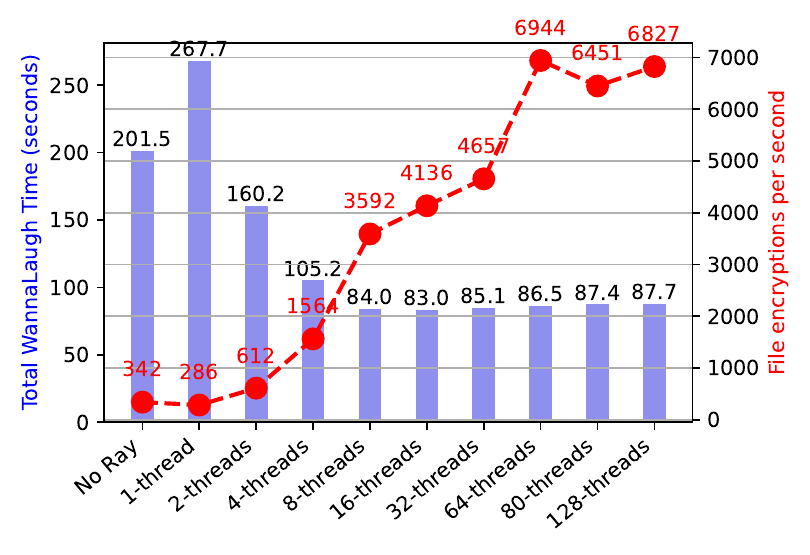}
    \caption{WannaLaugh Performance with Ray on Linux (80cores, XFS-INTEL\_SSDPE2KX010T8, 60 Govdoc1 folders, 59,272 files, 44.7 GB).} 
  \label{fig:plot_ray}
\end{wrapfigure}

As shown in Fig.~\ref{fig:plot_ray}, the Y-axis on the left presents the total time required by WannaLaugh to encrypt all files and write all corresponding AES keys to a file. The Y-axis on the right displays the number of encryptions per second, excluding the time taken to write the AES keys to a file. 

The baseline encryption process, mirroring a traditional ransomware's sequential file encryption, was completed in 201.5 seconds. When employing multi-threading with eight or more threads, this duration dropped significantly to 84 seconds. However, the performance gain plateaus beyond this point, primarily due to the time required to write the AES keys to a file becoming the dominant factor.

Remarkably, when assessing the number of encryptions per second, the performance boost is even more pronounced. Multi-threading achieves a staggering 20x increase in throughput compared to the baseline. Notably, the encryption of all 58,213 files was completed in just 15 seconds when we excluded the time to write the final AES keys and initialize the Ray cluster.

\subsection{Convergence Performance of Optimization Algorithms}

Starting from an initial random state, we allowed each algorithm to proceed for 1000 iterations. For every iteration, we performed an inference on a pre-trained model to get the F1 score. We then computed the minimum, maximum, and average values of the fitness cost across parallel solutions at each iteration, which also correspond to different parameter names, such as population for Genetic Algorithms or parallel executions for Random Walk.

In the context of our work, "fitness" or "fitness cost" refers to a metric that quantifies the quality of the generated trace, relative to a target trace. Table~\ref{tab:chosen_ransomware} shows the ransomware strains used for collecting target traces in this work, with their associated encryption schemes. For each of the four algorithms used, we adopt a common fitness measure to ensure comparability and interpretability across the methods. The fitness cost is calculated as a distance metric between the trace produced by the algorithm at each epoch and the actual ransomware trace. The nature of this distance calculation depends on the characteristics of the IO traces, taking into account the F1-score of the inference of the generated trace.

\begin{table}[h]
\centering
\scriptsize
        \caption{Selected Ransomware and SHA256 }
        \begin{tabular}{|p{0.1\linewidth} | p{0.30\linewidth}|p{0.21\linewidth}|} 
         \hline
         \textbf{Ransomware} & \textbf{SHA256} & \textbf{Algorithm}\\  
         \hline\hline
         \multirow{2}*{Black Basta} & \scriptsize 5d2204f3a20e163120f52a2e3595db19890050 b2faa96c6cba6b094b0a52b0aa & \scriptsize  \multirow{2}*{ChaCha20 \cite{Black-basta-trendmicro}}\\
         \hline
         \multirow{2}*{Lockbit 2.0} & \scriptsize 1adf694880ec194a166905cf1756cdb48ec9d9 8b6faaf7cd5d756c97bce93844 & \scriptsize \multirow{2}*{AES \cite{lockbitencryption}}\\
         \hline
         \multirow{2}*{Lockfile} & \scriptsize bf315c9c064b887ee3276e1342d43637d8c0e0 67260946db45942f39b970d7ce & \scriptsize \multirow{2}*{Intermittent encryption \cite{X-Force_Lockfile}}\\
         \hline
         \multirow{2}*{WannaCry} & \scriptsize 6e51fa23db7d2f83f3d380d6d465e32621793d 8c50e6bde255d996c25288c044 & \scriptsize \multirow{2}*{RSA and AES \cite{wannacryencryption}}\\
         \hline
         \multirow{2}*{Conti} & \scriptsize e22ff5594ab357da993651bf7decf035400d42b f37cca943495e1ac9bd7721c4 & \scriptsize \multirow{2}*{AES-256 \cite{decodingconti}}\\
         \hline
         WannaLaugh v0.1.46 & \scriptsize 6425e00a41247ffda3ec0155d87eaf7f375a10e ccd2ecb0004f8f8a17d74b1ce & \scriptsize \multirow{2}*{All the above}\\
         [1ex] 
         \hline
        \end{tabular}
\label{tab:chosen_ransomware}
\end{table}

Thus, a lower fitness cost indicates a higher resemblance between the generated trace and the actual ransomware trace, signifying a better "fit."  Our approach to fitness thus allows a direct comparison between the different search algorithms, despite their distinct internal mechanics, by providing a unified measure of how well they generate ransomware-like or benign-like traces.

\begin{figure}[h]
    \includegraphics[width=0.7\linewidth]{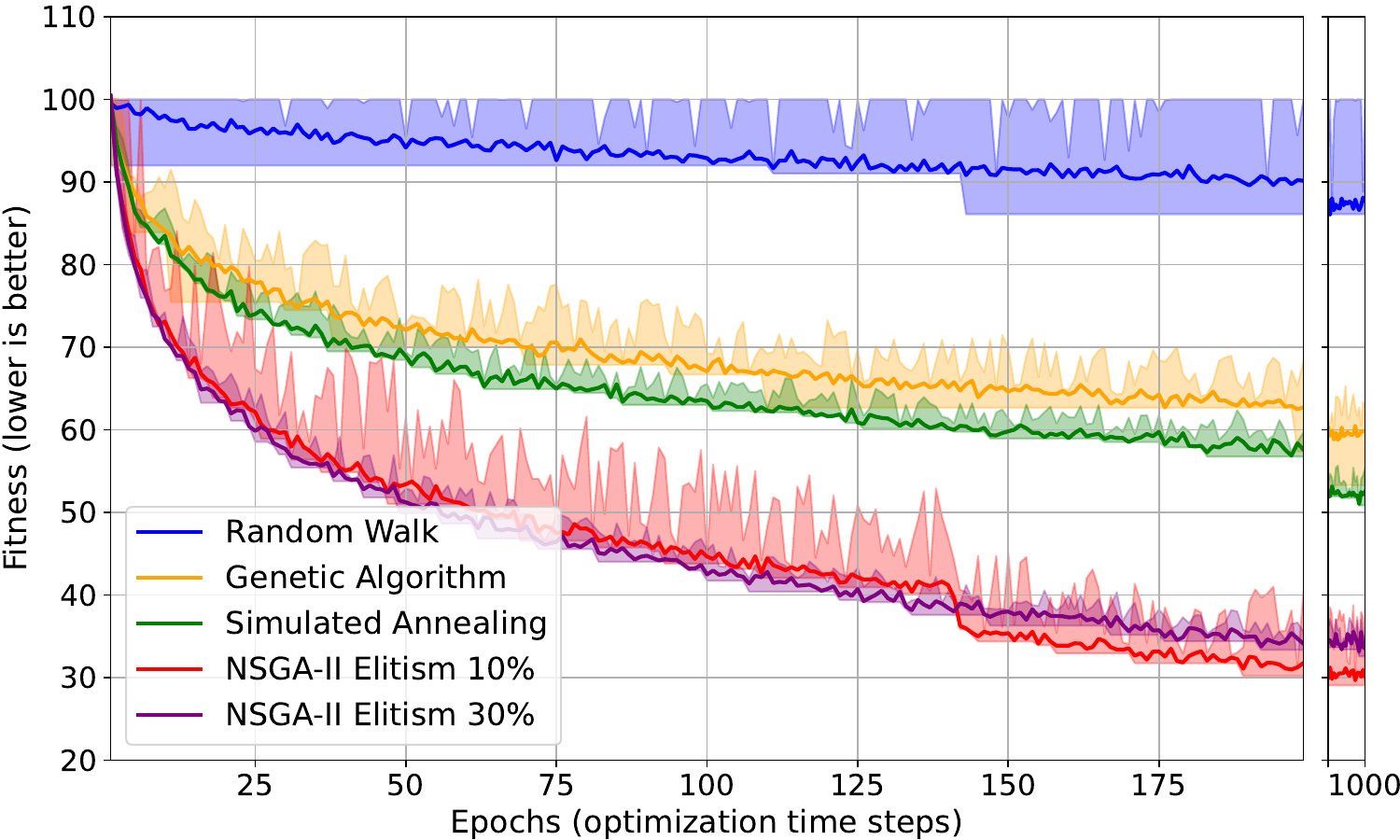}
    \caption{Optimizing imitating malicious strains with different heuristics.}
    \label{fig:optimization_algorithms_comparison}
    \vspace{-10pt}
\end{figure}

In Fig.~\ref{fig:optimization_algorithms_comparison}, we plot the fitness cost as a function of the number of epochs (generations/iterations/steps) with lower values being more desirable.  The dark line corresponds to the mean fitness cost for the parallel solutions (e.g., population) of each algorithm, while the light area indicates other solutions that were better or worse than the mean for a specific epoch.

While this figure presents the algorithm's iterative refinement process, it's important to understand the implications in the context of generating IO traces. Indeed, a trace can technically be generated using parameters identified by the search algorithms at any epoch, even at epoch 0. However, the overall accuracy and resemblance of the generated trace to the real one improve as the epochs progress, given the optimization nature of these algorithms. Therefore, the number of epochs can be seen as an indirect measure of the quality of the generated trace. It's not a direct indicator of real-time performance or end-to-end latency for generating a trace, but it provides valuable insight into the iterative refinement and optimization process of our approach.

For solutions with a fitness cost below the mean, we use a Pareto-style approach: if a solution at epoch $i$ has a lower fitness cost than its predecessor, 
it is considered for the next epoch. Otherwise, the solution from epoch $i-1$ is plotted as it represents a better outcome. This approach results in a Pareto-like representation of the graph for the solutions below their mean. NGSA-II outperforms RW, while GGA and SA exhibit intermediate convergence rates. After epoch 200, all algorithms except NSGA-II show negligible convergence with poor F1 scores below 0.5. In contrast, NSGA-II achieves a more promising F1 score of 0.71 and a faster convergence rate, highlighting its potential as a better optimization algorithm.

In addition, we conducted two experiments with NSGA-II, employing different levels of elitism: 10\% and 30\%~\cite{KUMAR201715}. Our analysis reveals that 30\% elitism initially demonstrated a faster convergence rate up to epoch 141. This can be attributed to the retention of a larger proportion of high-quality candidates within the solution pool. However, beyond epoch 141, using 10\% elitism outperforms the 30\% case, as it facilitated broader exploration of the design space. By utilizing 90\% of the solution pool for more diverse, random solutions at every iteration, the algorithm with 10\% elitism managed to avoid local optima, without limiting its search capability.

\subsection{Ability of WannaLaugh to Mimic Traces}
\subsubsection{Examining two versions of WannaLaugh}
\label{ssec:mimic_traces}

\begin{figure}[htbp]
    \centering
    \begin{subfigure}[b]{\linewidth}
        \centering
        \includegraphics[width=0.7\linewidth]{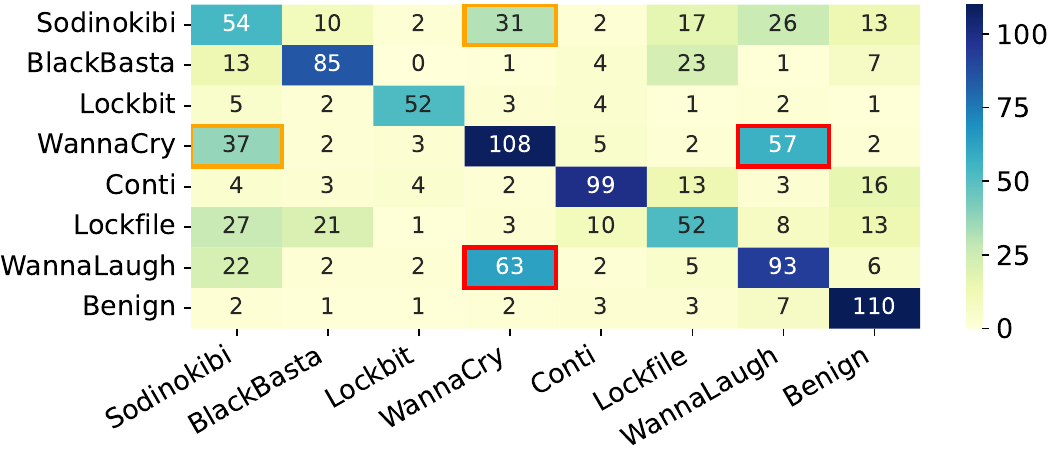}
        \caption{WannaLaugh-W (configured to resemble WannaCry).}
        \label{fig:sub1}
    \end{subfigure}
    \hfill
    \vspace{5pt}
    \begin{subfigure}[b]{\linewidth}
        \centering
        \includegraphics[width=0.7\linewidth]{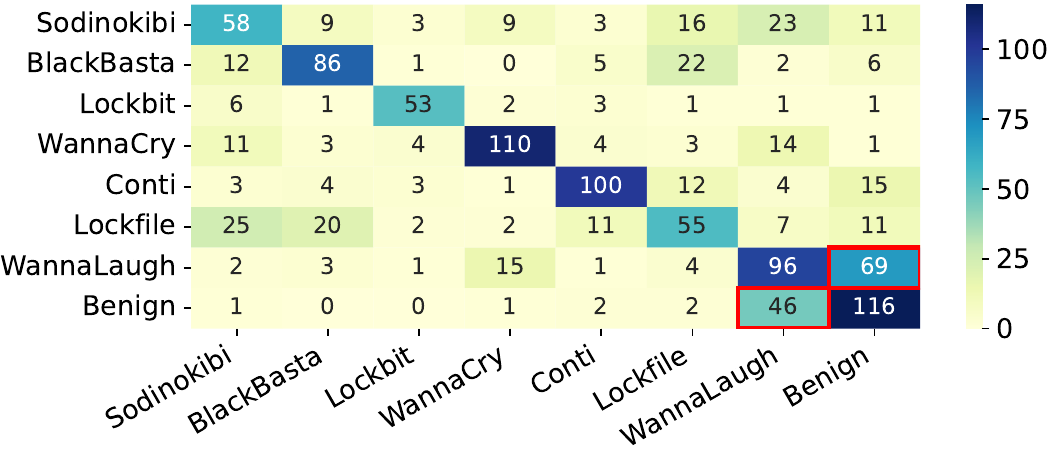}
        \caption{WannaLaugh-N (configured to not resemble any ransomware).}
        \label{fig:sub2}
    \end{subfigure}
    \caption{Confusion matrix of six ransomware, one benign workload and WannaLaugh.}
    \label{fig:heatmap}
\end{figure}

We continue our evaluation by using two versions of WannaLaugh using the optimal parameters from the previous analysis, namely, (1) WannaLaugh-W, which is designed to resemble a known ransomware strain, such as WannaCry, and (2) WannaLaugh-N, which is designed not to resemble any strain from the dataset, but to form a new potential strain and compare each of them with all six ransomware strains and one benign workload in the confusion matrices shown in Fig.~\ref{fig:heatmap}.

Each cell in the confusion matrix represents the intersection of the predicted class (column) and the true class (row) for a given instance. The diagonal cells (top left to bottom right), represent correct predictions. 
The off-diagonal cells signify misclassifications, where the model incorrectly predicted the class. In the case of the WannaLaugh-W and WannaLaugh-N, the high values in specific off-diagonal cells indicate the intended behavior of the respective ransomware versions, either resembling a known ransomware strain (red boxes of Fig.~\ref{fig:heatmap}(a)) or being classified as a benign workload (red boxes of Fig.~\ref{fig:heatmap}(b)).

The confusion matrices confirm our expectations. For WannaLaugh-W, we observe high values at the intersection of WannaLaugh-W and the WannaCry ransomware which indicates successful generation of traces that closely resemble known ransomware strains. For WannaLaugh-N, we observe high values in the cells corresponding to the intersection of WannaLaugh-N and benign workloads. This outcome demonstrates that we effectively created traces that can be classified as benign workloads, potentially evading detection by ransomware identification tools.

\subsubsection{Overall analysis of all WannaLaugh versions}

Building upon the time series data analysis presented above, Table~\ref{tab:comparison} offers a comprehensive summary of these results in the form of average match percentages for all collected traces. Every trace has a length of 2700s and the number of samples from each class is balanced. This representation provides an overarching view of Wannalaugh's ability to mimic a variety of different workloads under various configurations.

Each row in the table displays the average percentage to which Wannalaugh, when configured to emulate a specific workload, matches each of the six ransomware strains and one benign workload. The values are derived from a comprehensive analysis of the IO trace time series data for each configuration. For example, when Wannalaugh is set to replicate the 'WannaCry' ransomware, it corresponds to the IO trace of the actual 'WannaCry' ransomware with an average of 57.2\% accuracy. It is worth noting that these percentages significantly drop when the comparison involves different workloads, thereby demonstrating Wannalaugh's ability to differentiate between various ransomware strains.

\begin{table}
\centering
\caption{Evaluation of WannaLaugh mimicking capability (\%)}
\label{tab:comparison}
\begin{tabular}{l|c|c|c|c|c|c|c}
 & \rotatebox{70}{Sodinokibi} & \rotatebox{70}{BlackBasta} & \rotatebox{70}{Lockbit} & \rotatebox{70}{WannaCry} &  \rotatebox{70}{Conti} & \rotatebox{70}{Lockfile} & \rotatebox{70}{Benign} \\
\hline
\hline
WannaLaugh-Sodinokibi & \textbf{51.3} & 9.76 & 7.86 & 13.1 & 10.5 & 13.5 & 8.43 \\
\hline
WannaLaugh-BlackBasta & 25.9 & \textbf{52.7} & 7.76 & 9.23 & 7.21 & 8.32 & 11.7 \\
\hline
WannaLaugh-Lockbit & 9.31 & 8.72 & \textbf{23.4} & 7.64 & 9.28 & 10.4 & 7.23 \\
\hline
WannaLaugh-WannaCry & 20.0 & 1.8 & 1.8 & \textbf{57.2} & 1.8 & 4.5 & 5.4 \\
\hline
WannaLaugh-Conti & 14.7 & 8.94 & 11.6 & 9.89 & \textbf{65.1} & 12.4 & 13.1 \\
\hline
WannaLaugh-Lockfile & 11.3 & 11.5 & 10.5 & 14.7 & 10.9 & \textbf{62.3} & 13.0 \\
\hline
WannaLaugh-Benign & 1.7 & 2.5 & 0.8 & 12.9 & 0.8 & 3.4 & \textbf{59.4} \\
\hline
\end{tabular}
\vspace{-10pt}
\end{table}

The bold figures along the diagonal of the table denote instances where Wannalaugh is configured to match the same workload that it is compared with. In reviewing these findings we note an encouraging trend. The diagonal percentages, reaching up to 65.1\%, reflect Wannalaugh's promising ability to mimic various workloads with a degree of accuracy. This demonstration of performance points to the effectiveness of the current configurations.

However, these percentages are far from perfect, indicating that there is room for enhancement and fine-tuning of Wannalaugh's configuration parameters. It is important to underline that we do not strive for 100\% emulation accuracy. Given the inherent non-deterministic nature of IO traces and the influence of factors such as operating system noise, achieving a perfect match is not only unrealistic but also unnecessary. Moreover, an accuracy of 100\% could potentially signal a risk of overfitting, as it might indicate the model is too specifically tailored to the training data.

Still, the difference between the current maximum of approximately 65.1\% and a perfect match does signal a wide margin for further refinement of Wannalaugh's parameters. This study's results serve as a promising baseline upon which future work can build, driving the continual improvement of Wannalaugh's emulation capabilities.

\section{Future Work}

We are currently investigating the following extensions for the WannaLaugh ransomware emulator: (1) the integration of more advanced search strategies, such as reinforcement learning to guide the selection of configuration options more efficiently~\cite{zhang2021dual}. This would allow for faster convergence to optimal or near-optimal settings. 
(2) The support for classification against a broader range of popular ransomware types, enabling the development of more general ML-based detection models which would improve the emulator's ability to cover a wider variety of ransomware behaviors and encryption patterns. 
(3) The effectiveness in detecting ransomware when different disk setups, RAID systems, or filesystems are involved. Exploring this aspect would ensure the emulator remains versatile and adaptable to various system configurations, further broadening its applicability and practicality in real-world scenarios.
(4) We further investigate the integration of trace generators, such as SDGen~\cite{gracia-tinedo:2015}, Tracie~\cite{sfakianakis:2021} or Sysbench~\cite{sysbench}, to support more benign types of workloads.

Another intriguing possibility for ransomware trace generation is the utilization of AI-driven generative models such as Generative Adversarial Networks (GANs)~\cite{chen2018towards}. However, the lack of constrained interpretability nature, the need for extensive training data, and the excessive computational demands can pose significant hurdles. Before exploring these complex methods, it's beneficial to establish the viability of the approach using simpler machine learning models. Hence, we've chosen a more supervised and less resource-intensive approach, while also ensuring control over the generation process. The exploration of AI and generative models for this task remains a promising avenue for future research.

Since WannaLaugh is open-sourced\footnote{WannaLaugh is going to be open-sourced on Github and PyPi soon.}, the cybersecurity community can actively contribute to these extensions and improvements. Involving the community can ensure that the emulator evolves alongside emerging threats, helping researchers develop more robust and effective solutions against ransomware and other cyber threats. These continuous enhancements will solidify WannaLaugh as a valuable tool. 

Last but not least we would like to mention that WannaLaugh significantly helped us to train ML models for an enterprise storage system using CSD capabilities in SSDs to detect ransomware in real-time. While it would be very interesting to depict such an architecture and its constraints and demonstrate the accuracy, generalizability, and limitations of such ML models we looked into, this topic is clearly out of the scope of this paper and will be addressed in the near future.

\section{Conclusion}

This paper introduces WannaLaugh, a configurable open-source ransomware emulator designed to generate IO traces in a safe and controlled environment. By leveraging heuristic algorithms to explore the emulator's extensive design space, WannaLaugh can produce realistic IO traces that either resemble known ransomware behaviors or not. This novel tool enables researchers to develop, evaluate, and refine ML-based ransomware detection techniques, addressing the ever-evolving threat landscape. The highly customizable and extensible nature of WannaLaugh ensures its continued relevance and applicability in the rapidly changing field of cybersecurity, paving the way for more robust and efficient ransomware detection and prevention strategies.

\bibliographystyle{unsrtnat}

\bibliography{wannalaugh-sigmetrics2024.bib}

\end{document}